\documentclass[12pt]{elsarticle}
\begin{document}

\title{Domenico Pacini, the forgotten pioneer of the discovery of cosmic rays}

%\title{Domenico Pacini, the forgotten pioneer of the discovery of cosmic rays}  

\author{A De Angelis$^{1},$ N. Giglietto$^{2}$ and S. Stramaglia$^{2}$}
\address{$^1$	Dipartimento di Fisica dell'Universit\`a di Udine and INFN, Via delle Scienze, 33100 Udine, Italy; INAF Trieste, Italy; LIP/IST, Lisboa, Portugal}
\address{$^2$	Dipartimento Interateneo di Fisica di Bari and INFN, Via Orabona, 70126 Bari, Italy}

\begin{abstract}
About a century ago, cosmic rays were identified as being a source of radiation on Earth. 
The proof came from two independent experiments. The Italian physicist Domenico Pacini observed the radiation 
strength to decrease when going from the surface to a few meters underwater (both in a lake and in a
sea). At about the same time, in a balloon flight, the Austrian Victor Hess  found the ionization 
rate to increase with height. The present article attempts to give an unbiased historical account 
of the discovery of cosmic rays -- and in doing so it will duly account for Pacini's pioneering 
work, which involved a technique that was complementary to, and independent from, Hess'. Personal 
stories, and the pre- and post-war historical context, led Pacini's work to slip into oblivion. 
\end{abstract}
%\pacs{98.70.Sa; 01.65.+g}

\maketitle

\section{Introduction}

It is generally well known, since Faraday's early observations, that electroscopes 
spontaneously discharge. This phenomenon remained unexplained until the beginning 
of the XX century: its explanation paved the way to one of mankind's revolutionary 
scientific discoveries: cosmic rays. 

Since the early XX century cosmic rays were used to probe and understand the  
constituents of matter. Indeed many early discoveries in particle physics 
(antimatter, mesons, muons, baryons, ...) were made while studying cosmic rays. 
Cosmic rays are still being used in the framework of fundamental physics, as well 
as to investigate astrophysical properties of their sources.

In 1896 the French physicist Henri Becquerel discovered the instability of some chemical 
elements. Some years later Marie and Pierre Curie discovered that Radium showed that same 
behavior: such transmutation processes were then called ``radioactive decays''. In the 
presence of a radioactive material, a charged electroscope promptly discharges. It was 
concluded then that some elements were able to emit charged particles, that in turn were 
responsible for discharging the electroscope. An electroscope's discharge rate was then 
used to gauge the level of radioactivity. 

The spontaneous discharge observed in electroscopes made it evident that in insulated 
environments, too, a background radiation did exist. The obvious questions concerned the 
nature of such radiation, and whether it was of terrestrial or extra-terrestrial origin. 
It was generally believed that its origin was likely related to radioactive materials, 
hence its terrestrial origin was a commonplace assumption. An experimental proof, however, 
seemed hard to achieve.

At the very beginning of the XX century, several scientists made experiments about 
penetrating radiation, trying to understand its origin and nature\footnote{Cline in 1910 summarizes~\cite{clin} the status of the art:
experiments were mainly oriented
to measure the daily variations or seasonal variations. Cline cited the work by the Italian Domenico Pacini~\cite{pac0} about the daily variations of the radiation measured on the
sea at Sestola, in Italy. Pacini's measurement was remarked in Cline's paper as a
first evidence of the atmosphere being the main responsable of the penetrating radiation,
excluding the Sun as the main origin.}.

Around 1910, the Austrian Victor Hess and  the Italian Domenico 
Pacini simultaneously and independently carried out two different, ingenious, and 
complementary research lines that would eventually clarify the origin of the yet 
mysterious ionizing background radiation.

Pacini made several measurements to establish the variations of an Ebert electroscope's 
discharge rate as a function of the environment: he placed the electroscope on the 
ground, on the sea a few km off the coast, and a few meters underwater. He reported 
those measurements, the ensuing results, and their interpretation 
in a note titled ``La radiazione penetrante alla superficie ed in seno alle acque''  \emph{(``Penetrating radiation at the surface of and in water'')}
\cite{pacini12}. In that paper Pacini wrote: ``Observations carried out on the sea 
during the year 1910~\cite{pacini11} led me to conclude that a significant proportion 
of the pervasive radiation that is found in air had an origin that was independent of 
direct action of the active substances in the upper layers of the Earth's surface.''
What was he lacking, at that time, before he could reach a firm conclusion about the 
extraterrestrial origin of the ionizing background radiation? Only in 1911 did Pacini 
develop his experimental technique for underwater measurements, that allowed him to 
measure a significant decrease in the discharge rate when the electroscope was placed 
underwater. {``The apparatus ...  was enclosed in a copper box so that it could immerse 
in depth.  ... From June 24 to June 30 observations were performed with the instrument 
at the surface, and with the instrument immersed in water, at a depth of 3 meters. ...
[It] appears from the results of the work described in this Note that a sizable cause of 
ionization exists in the atmosphere, originating from penetrating radiation, independent 
of the direct action of radioactive substances in the soil."}

Who was Domenico Pacini and, while Victor Hess~\cite{hes} is honored as the discoverer 
of cosmic rays, why did Pacini's contemporary (or even earlier) discovery go unnoticed and 
was soon forgotten (notably in Italy)? Personal stories and historical events contributed 
to this outcome.

\begin{figure}
\begin{center}
\includegraphics[width=0.5\textwidth]{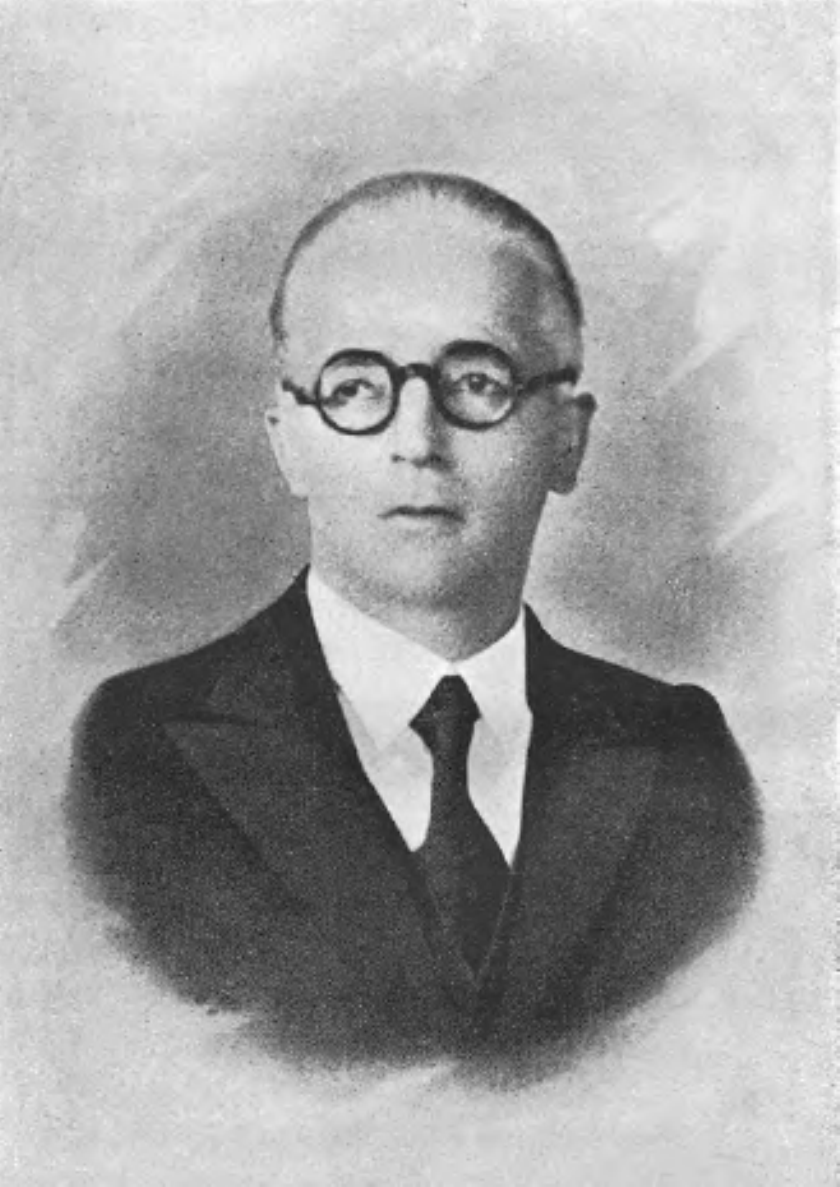}\label{fig:pacini}
\end{center}
\caption{Domenico Pacini.}
\end{figure}

Domenico Pacini (Figure \ref{fig:pacini}) was born on February 20, 1878, in Marino, near Rome. He graduated in 
Physics in 1902 at the Faculty of Sciences of Rome University. There, for the 
next three years, he worked as an assistant to Professor Pietro Blaserna while also 
studying electric conductivity in gaseous media under the supervision of Alfonso 
Sella. In 1904 he set out to study the infamous N-rays: he performed an 
experiment, the (null) results of which were communicated in a letter to Nature~\cite{nrays}
as ``careful
experiments made ... with the object of observing the
effects of n-rays described by M. Blondlot and other investigators.''
Though ``observations were made under very
favourable conditions,'' he ``was unable to detect any increase
of luminosity of a phosphorescent screen caused by unknown
rays from strained or tempered steel, an Auer lamp, a
Nernst lamp, sound vibrations, or a magnetic field, though
various French observers have affirmed that in each of these
cases N-rays are emitted which produce an effect upon the
screen''. In 1906 
Pacini was appointed assistant at Italy's Central Bureau of Meteorology and Geodynamics, 
heading the department that was in charge of studying thunderstorms and electric phenomena 
in the atmosphere. (Most of the department's experimental work was carried out near 
Castelfranco Veneto, near Padova.) 
Pacini's held that position until 1927, when he was upgraded to Principal Geophysicist.
After several more years of work as an assistant professor in Rome, finally in 1928 he 
was appointed full professor of Experimental Physics at the University of Bari, where he was incharged of setting up the studies of  Physics within the Faculty of Medicine. While in Bari, 
his research interests mainly focused on the diffusion processes of light in the atmosphere. 
Domenico Pacini died of pneumonia in Rome on May 23, 1934, shortly after his marriage.

\begin{figure}
\begin{center}
\includegraphics[width=0.7\textwidth]{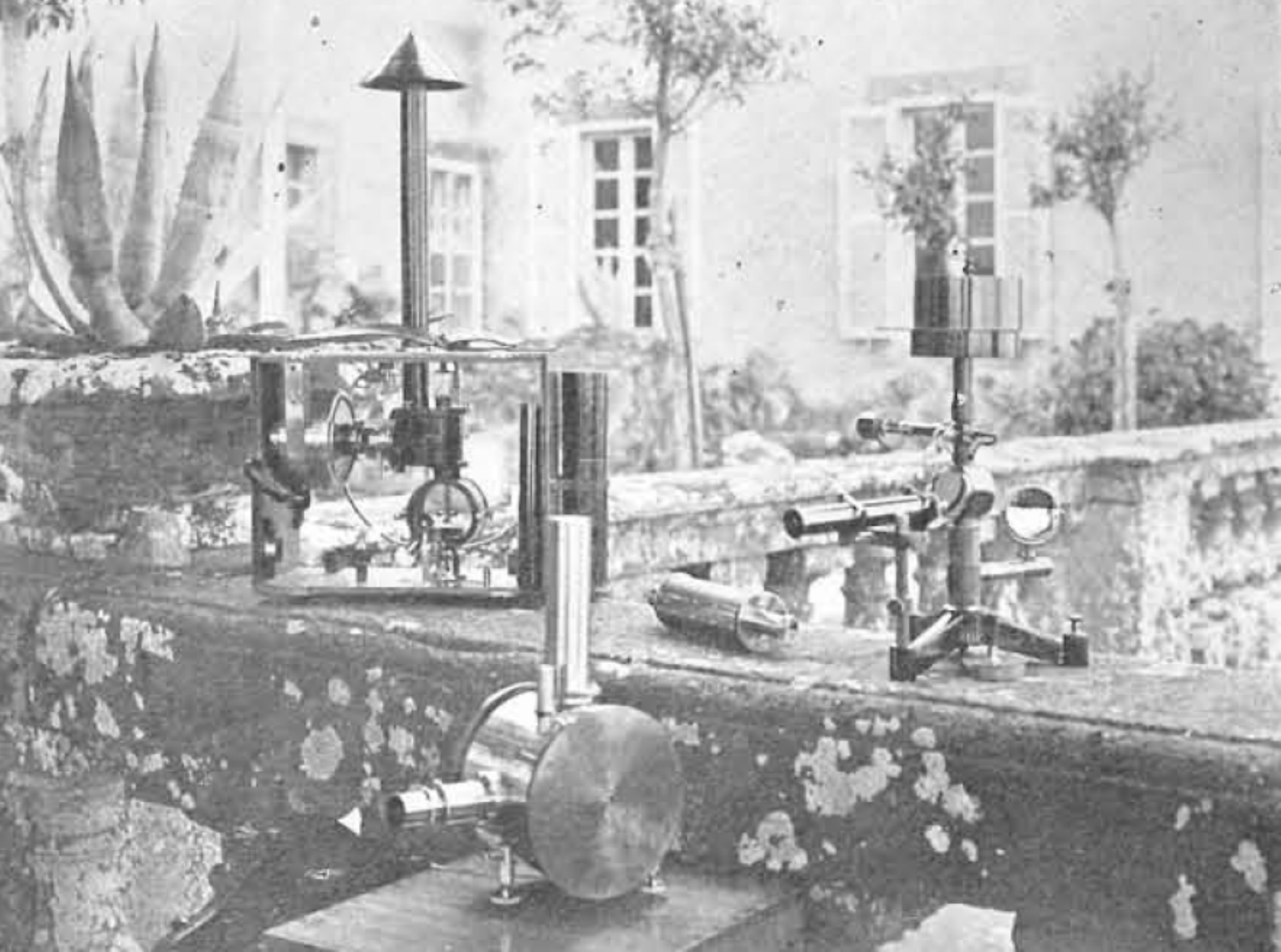}\label{fig:elettroscopi}
\end{center}
\caption{Three electroscopes actually used by Pacini for his measurements. In foreground, the Ebert's electroscope.}
\end{figure}

The long way that led Pacini to the hypothesis of cosmic rays started from his studies 
on electric conductivity in gaseous media that he performed at the University of Rome 
during the early years of the XX century. While working at the Central Bureau of 
Meteorology, he became interested in the problem of the ionization of air. During 
1907--1912, he performed several measurements on the air's conductivity on the ground 
(at different elevations, including at sea level), on the sea, and (later in 1911) 
underwater~\cite{pacini11,pacini12,paciniot}. Those measurements, performed with electroscopes (Figure \ref{fig:elettroscopi}),  were aimed at checking 
whether the radioactivity within the Earth's crust was sufficient to explain the 
ionization effects (about 13 ions per second per cubic centimeter of air) that had been 
measured on the Earth's surface. Pacini concluded that the Earth's radioactivity alone 
was not sufficient to explain the observations. 

\begin{figure}
\begin{center}
\includegraphics[width=0.8\textwidth]{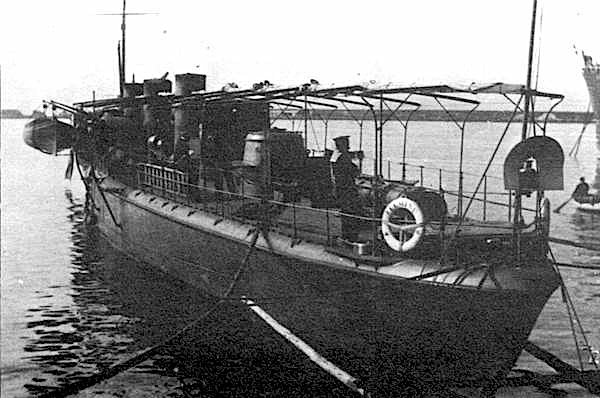}\label{fig:nave}
\end{center}
\caption{The cacciatorpediniere ``Fulmine'', used by Pacini for his measurements on the sea.}
\end{figure}

Pacini made several measurements on an Italian Navy ship (the {\emph{cacciatorpediniere}} 
``Fulmine'', Figure \ref{fig:nave}). First he concluded that the ionization above the sea, at sea level, and far 
from the coast, was consistent with that measured at ground level on land. From 1910 on, 
Pacini proposed a new experimental technique that proved very successful and was to became 
important for later developments of physics: he measured the radiation intensity in water, 
at a depth of 3 meters, in the Genoa Gulf and in the Bracciano Lake (near Rome), proving 
that the radiation was significantly smaller underwater than on the ground. To explain his 
results, that marked the beginning of the underground/underwater technique for cosmic-ray studies 
(that has been implemented so many times up to this day), Pacini proposed the existence of 
a radiation of extraterrestrial origin -- later to be called ``cosmic rays''. A few months 
later Victor Hess, using balloon flights, confirmed those results with measurements that 
eventually earned him the Nobel Prize for Physics in 1936, two years after Pacini's death.  

In his paper %(Figure~\ref{fig:nuovocim})
 containing the interpretation of such results~\cite{pacini12}, published a 
few months before the paper by Hess, Pacini was evidently aware of Hess' results as he did 
quote them correctly. Some excerpts from mail exchanges that occurred between the two scientists in 1920 
are very illuminating on the issue of the priority in discovering cosmic rays. (Such 
exchanges were reported in the Pacini commemoration held 
in Bari in 1935.) On March 6, 1920, Pacini wrote to Hess: ``...I had the opportunity to study 
some of your papers about electrical-atmospherical phenomena that you submitted to the Principal 
Director of the Central Bureau of Meteorology and Geodynamics. I was already aware of some of these 
works from summaries that had been reported to me during the war. [But] the paper entitled ``Die Frage 
der durchdring. Strahlung ausserterrestrischen Ursprunges'' \emph{(``The problem of penetrating 
radiation of extraterrestrial origin'')} was unknown to me. While I have to congratulate 
you on the clarity in which this important matter is explained, I have to remark, unfortunately, 
that the Italian measurements and observations, which take priority as far as the conclusions 
that you, Gockel and  Kolh\"orster draw, are missing; and I am so sorry about this, because in my own 
publications I never forgot to mention and cite anyone...''. The answer by Hess, dated March 17, 1920, was: ``Dear Mr. Professor, your very valuable 
letter dated March 6 was to me particularly precious because it gave me the opportunity to 
re-establish our links that unfortunately were severed during the war. I could have contacted you before, 
but unfortunately I did not know your address. My short paper ``Die Frage der durchdring. 
Strahlung ausserterrestrischen Ursprunges'' is a report of a public conference, and therefore 
has no claim of completeness. Since it reported the first balloon measurements, I did not provide 
an in-depth explanation of your sea measurements, which are well known to me. Therefore please  
excuse me for my unkind omission, that was truly far from my aim ...''. On April 12, 1920, 
Pacini in turn replied to Hess: ``... [W]hat you say about the measurements on the 
penetrating radiation performed on balloon is correct; however the paper ``Die Frage der durchdring. 
Strahlung ausserterrestrischen Ursprunges'' lingers quite a bit on measurements of the attenuation 
of this radiation made before your balloon flights, and several 
authors are cited whereas I do not see any reference to my relevant measurements (on the 
same matter) performed underwater in the sea and in the Bracciano Lake, that led me to  
the same conclusions that the balloon flights have later confirmed.''

%\begin{figure}
%\begin{center}
%\includegraphics[width=0.5\textwidth]{CoverPacini12.pdf}\label{fig:nuovocim}
%\end{center}
%\caption{The cover of the ``Nuovo Cimento'' issue in which the conclusive measurements by Pacini have been published.}
%\end{figure}

Edoardo Amaldi had no doubt that Domenico Pacini was indeed the discoverer of cosmic rays. 
This is  reported in a letter that E.~Amaldi wrote on July 14, 1941, to the Director of the 
Physics Institute of Rome University, Antonino Lo Surdo (the letter belongs to the 
Amaldi Archive at the ``La Sapienza'' University of Rome~\cite{rob}). E.~Amaldi's letter 
was motivated by an article that had earlier appeared on a local newspaper, in which it was 
stated that nuclear physics and cosmic ray physics were ``Jewish sciences''. Here is the relevant 
quote from E.~Amaldi's answer: ``... this statement appear so strange to anyone who knows, as 
you certainly do, that the Italian Domenico Pacini was the discoverer of the Cosmic Rays and 
only afterwards did the German Hess, Kolh\"orster, etc. follow ...".
  
The route to cosmic rays, opened by Pacini, continued in the following years and, after 
the second world war, Italy took up a leading role in the field. Back in those years most of the research 
was performed at laboratories placed on mountain tops. In Italy the most important one was 
located at Testa Grigia on the Plateau Rosa, at an altitude of 3500 meters on the Matterhorn.  
A team from the University of Padova established an observatory at the Passo Fedaja on the Dolomites near 
Belluno, at an elevation of approximately 2000 meters: thanks to the Societ\`a Adriatica di 
Elettricit\`a (SADE), the observatory had enough electric power to run a large magnet for 
experiments that used counters and cloud chambers. 

The complementary work by Pacini and Hess were seminal in starting, respectively, underwater 
and upper-atmosphere/space studies of cosmic rays. 

Is it possible to say now, as Edoardo Amaldi did once, that ``the Italian Domenico Pacini was the 
discoverer of the Cosmic Rays [and was followed] by the Germans Hess, Kolh\"orster etc.''? 
A great discovery is in general the result of joint efforts by many people. It is certainly 
true that Pacini wrote, in as early as 1910 \cite{pacini11}, that the action of active substances 
in the soil was not sufficient to explain the observed properties of the penetrating radiation; 
and that he was the first to publish such a statement. However, a whole community of researchers 
was already involved in that field. Pacini's work was carried out in difficult conditions because 
of lack of resources available to him, because of lack of scientific freedom during the crucial 
years when he was working at the Central Bureau of Meteorology and Geodynamics, and finally  
because of the substantial indifference his work was met with by the Italian academic world -- a 
fact that still today is sadly evident to anyone who treads non-mainstream scientific paths. 

We conclude by remarking that, as described in the Bulletin of the Societ\`a Aeronautica Italiana, 
at the beginning of the XX century an Italian balloon, named after Jules Verne, did reach a height 
of about 5000 meters. This proves that in Italy, too, a balloon-borne experiment would have been 
technically feasible in those years. An open-minded national research agency -- 
such as it did not exist in Italy in those days -- can really be the key to success in experimental 
Physics. This may be a lesson Domenico Pacini has taught us, for the present time as well as for 
future times.

\section*{Acknowledgements}
%\ack

We are grateful to the University of Bari, and in particular to Professor A.~Garuccio, 
for supporting the research of documents regarding Domenico Pacini; to the Dipartimento 
Interateneo di Fisica of Bari for organizing the Domenico Pacini memorial day that 
was held in Bari on April 17, 2007; to Professors F.~Guerra and N.~Robotti for uncovering relevant 
material in the Amaldi Archive at Rome's ``La Sapienza'' University and in the Bracciano 
museum; to Roberto Garra for uncovering material in ``Collegio Romano''; to Comandante E.~Bagnasco, A.~Lombardi from Associazione Culturale Italia, C.~D'Adamo from Regia Marina Italiana for historical pictures and information on the ships by the Italian Navy; and to Professors L.~Guerriero, E.~Menichetti, P.~Spinelli, L.~Cifarelli, P.~Carlson and M.~Persic for help, support, discussions and suggestions.

%\section*{References}

\end{document}